\documentclass[aps,prb,twocolumn,showpacs]{revtex4}

\usepackage{graphicx}
\usepackage{amssymb}

\begin{document}

\title{Pressure Induced Charge Disproportionation in LaMnO$_{3}$}

\author{G. Banach and W.M. Temmerman}
\affiliation{
Daresbury Laboratory, Daresbury, Warrington WA4 4AD, UK }

\date{\today}

\begin{abstract}
We present a total energy study as a function of volume in the cubic phase of LaMnO$_{3}$.
A charge disproportionated state into planes of Mn$^{3+}$O$_{2}$/Mn$^{4+}$O$_{2}$ was found. It is argued that the 
pressure driven localisation/delocalisation transition might go smoothly through  
a region of Mn$^{3+}$ and Mn$^{4+}$ coexistence. 
\end{abstract}


\maketitle


\section{Introduction}

The pressure-induced quenching of the Jahn-Teller distortion and metal-insulator transition in LaMnO$_{3}$ 
were recently studied \cite{loa} by synchrotron x-ray diffraction, optical spectroscopies, 
and transport measurements under pressures up to 40 GPa. 
This work stresses that the delocalisation of electron states is a key feature of LaMnO$_{3}$ 
in the pressure range of 20-30 GPa. In this paper we present calculations which show 
that the delocalisation of electron states goes through a phase of charge disproportionation 
where Mn$^{3+}$ and Mn$^{4+}$ coexist.

Interest in perovskite manganites\cite{ramirez,imada,coey} was recently rekindled due to the observation of 
'colossal' negative magnetoresistance in La$_{1-x}$CaMnO$_{3}$.\cite{jin} 
The half-metallic character of La$_{2/3}$Sr$_{1/3}$MnO$_{3}$\cite{bowen} is an important factor 
which results in a magnetoresistance ratio of more than 1800\% at 4K. 
The phase diagram of La$_{1-x}$SrMnO$_{3}$ is complex,\cite{hemberger} 
revealing the competition between the double exchange and superexchange interactions, 
charge and orbital ordering instabilities and strong coupling to the lattice deformations. 
Furthermore it has been argued that local Jahn-Teller effects, 
such as random Jahn-Teller distortions of the MnO6 octahedra,\cite{dzero} 
and dynamical effects\cite{michaelis} are of importance to explain the magnetic properties of these materials. 

Recently Banach and Temmerman\cite{banach} showed that under Sr doping the delocalisation transition from 
the valence Mn$^{3+}$, as seen in LaMnO$_{3}$,\cite{rik} to the valence Mn$^{4+}$ is characterised by a transition region 
extending up to 20\% Sr where Mn$^{3+}$ and Mn$^{4+}$ coexist in a charge disproportionated state. 
These LSMO systems were cubic and therefore the Jahn-Teller distortion did not play a role in this. 
Similar results are also obtained by calculations of Korotin {\it et al.}\cite{korotin} 
who show for La$_{7/8}$Sr$_{1/8}$MnO$_{3}$ the occurrence of orbital order and insulating behaviour 
in the ferromagnetic state without any Jahn-Teller distortion.   

In this paper we study whether a delocalisation of Mn$^{3+}$ to Mn$^{4+}$ also occurs, under pressure,
 in the 
cubic phase of the parent compound LaMnO$_{3}$ and if such a pressure-induced transition is also 
characterised by a region of Mn$^{3+}$ and Mn$^{4+}$ coexistence.
Concerning the Jahn-Teller distortion, which is ignored in the present study, 
it was included in the study of the electronic properties of LaMnO$_{3}$ by Tyer {\it et al.}\cite{rik}.
It was found in that work that the Jahn-Teller interaction is the
dominant effect in realizing orbital order, which is not considered in this work.  

SIC-LSD calculations are performed for LaMnO$_{3}$ in the ferromagnetic (FM) and 
anti-ferromagnetic A (AF-A) - which is anti-ferromagnetic ordering in the (001) direction - 
structures at different volumes
to study the localization of the Mn d states. 
These calculations are performed for different configurations of localized states 
as a function of volume. In particular configurations where the 3d Mn t$_{2g}$ states 
are localised will be considered as well as 
3d Mn t$_{2g}$ and 3d Mn e$_{g}$ of 3z$^{2}$-r$^{2}$ symmetry. 
By doubling the unit cell in the z-direction we can also investigate 
in the AF-A structure and a Mn$^{3+}$/Mn$^{4+}$ charge ordered state with the same structure.

The organization of this paper is as follows: in the next section we explain
the SIC-LSD formalism and give some calculational details. The third section contains
our results and discussions and the fourth section summarizes the paper.

\section{Formalism and Calculational Details}

The basis of the SIC-LSD formalism is a self-interaction free total energy functional, 
\( E^{SIC} \), obtained by subtracting from the LSD total energy functional,
\( E^{LSD} \), a spurious self-interaction of each occupied electron state 
\( \psi _{\alpha } \)\cite{pedrew}, namely 
\begin{equation}
\label{eq1}
E^{SIC}=E^{LSD}-\sum _{\alpha }^{occ.}\delta _{\alpha }^{SIC}.
\end{equation}
 Here \( \alpha  \) numbers the occupied states and the self-interaction correction
for the state \( \alpha  \) is 
\begin{equation}
\delta _{\alpha }^{SIC}=U[n_{\alpha }]+E_{xc}^{LSD}[\bar{n}_{\alpha }],
\end{equation}
with \( U[n_{\alpha }] \) being the Hartree energy and \( E_{xc}^{LSD}[\bar{n}_{\alpha }] \)
the LSD exchange-correlation energy for the corresponding charge density \( n_{\alpha } \)
and spin density \( \bar{n}_{\alpha } \). 
The SIC-LSD approach can be viewed as an extension of LSD
in the sense that the self-interaction correction is only finite for spatially
localised states, while for Bloch-like single-particle states \( E^{SIC} \)
is equal to \( E^{LSD} \). Thus, the LSD minimum is also a local minimum of
\( E^{SIC} \). A question now arises, whether there exist other competitive
minima, corresponding to 
localised states, which could benefit
from the self-interaction term without loosing too much 
of the energy associated with band formation.
This is often the case for rather well localised electrons like the 3$d$
electrons in transition metal oxides. 
It follows from minimisation of Eq. (\ref{eq1}) that within the SIC-LSD 
approach
such localised electrons move in a different potential than the delocalized
valence electrons which respond to the effective LSD potential. For example,
in the case of manganese, three (Mn$^{4+}$) or four (Mn$^{3+}$)
Mn $d$ electrons move in the SIC potential, while all other electrons feel 
only the effective LSD potential. Thus, by including
an explicit energy contribution for an electron to localise, the ab-initio SIC-LSD
describes both localised and delocalized electrons on an equal footing, leading
to a greatly improved description of static Coulomb correlation effects over
the LSD approximation. 

In order to make the connection between valence and localisation more explicit
it is useful to define the nominal valence as
\[
N_{val}=Z-N_{core}-N_{SIC},
\]
where $Z$ is the atomic number (25 for Mn), $N_{core}$ is the number of core 
(and semi-core) electrons (18 for Mn), and $N_{SIC}$ is the number of localised, 
i.e., self-interaction corrected, states (either three or four for 
Mn$^{4+}$ and Mn$^{3+}$ respectively). Thus, in this formulation the valence is equal to the 
integer number of electrons available for band formation. 
To find the valence we 
assume various atomic configurations, consisting of different numbers of localised 
states, and minimise the SIC-LSD energy functional of Eq. (\ref{eq1}) with respect 
to the number of localised electrons. The SIC-LSD formalism is governed by the 
energetics due to the fact that for each orbital the SIC differentiates between 
the energy gain due to hybridisation of an orbital with the valence bands and the 
energy gain upon its localisation. Whichever wins determines if the
orbital is part of the valence band or not and in this manner
also leads to the evaluation of the valence of elements involved.

In the present work the SIC-LSD approach, has been implemented \cite{sic} 
within the linear muffin-tin-orbital (LMTO) atomic sphere approximation (ASA) band 
structure method, \cite{oka75} in the 
tight-binding representation. \cite{AJ84}

\section{Results and Discussions}

\begin{figure}
\includegraphics*[scale=0.3, angle=-90]{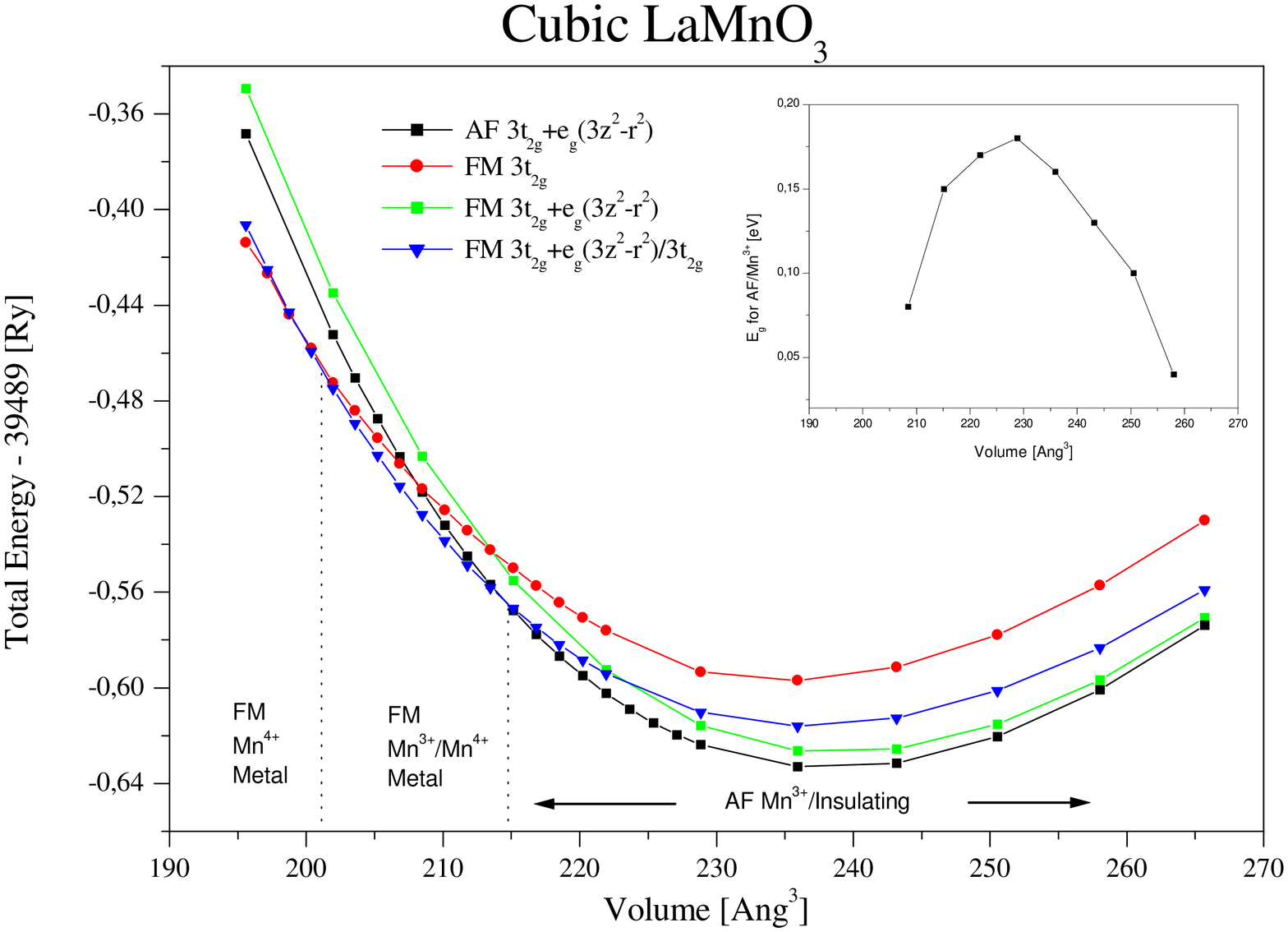}
\caption{\label{energyvsvol}
The total energy versus volume for cubic LaMnO$_{3}$
for four different scenarios of magnetic structure
and charge order. These are: anti-ferromagnetic A and trivalent Mn 
(with the three t$_{2g}$ electrons and one e$_{g}$ electron of $d_{3r^{2}-z^{2}}$ symmetry
localised) (black); ferromagnetic and tetravalent Mn (with the three t$_{2g}$ electrons localised) (red);
ferromagnetic and trivalent Mn (with the three t$_{2g}$ electrons 
and the e$_{g}$ electron of $d_{3r^{2}-z^{2}}$ symmetry localised) (green); 
the ferromagnetic Mn$^{3+}$/Mn$^{4+}$ charge disproportionated configuration (blue).
The inset shows the band gap versus volume for trivalent Mn in the AF-A structure.   
}
\end{figure}

In Fig. \ref{energyvsvol} we show the total energy versus volume for two different magnetic 
structures and three different states of charge order. These are the ferromagnetic and
anti-ferromagnetic (along the z-direction) magnetic orders. The charge order refers
to trivalent Mn$^{3+}$, tetravalent Mn$^{4+}$ and a disproportionated state of Mn$^{3+}$/Mn$^{4+}$.
This disproportionated state Mn$^{3+}$/Mn$^{4+}$ consists alternate MnO$_{2}$ planes in the
(001) direction.
The ground state is anti-ferromagnetic with Mn$^{3+}$ valence in the cubic structure. This result shows
that the Jahn-Teller distortion is not a necessary prerequisite for an insulating ground state.
Rather the valence of the Mn ion and the magnetic structure determine 
the insulating properties. Furthermore, the experimentally
observed AF-A magnetic order is obtained in the cubic structure, indicating that the Jahn-Teller
distortion does not influence the magnetic structure. Further inspection of Fig. \ref{energyvsvol} also
reveals that the energy scale associated with the magnetic structure is smaller than the one associated 
with the charge. In the vicinity of the energy minimum we find that the least unfavourable
configuration is trivalent (Mn$^{3+}$) in the ferromagnetic state, whilst the most
unfavourable is tetravalent (Mn$^{4+}$) in the ferromagnetic structure. 
The theoretical volume is in very good agreement with experiment: 238 \AA$^{3}$ versus an experimental
value of 244 \AA$^{3}$.

Reducing the volume, we obtain between 200 \AA$^{3}$  
and 215 \AA$^{3}$  a charge disproportionated  
Mn$^{3+}$/Mn$^{4+}$ state, in the ferromagnetic structure, 
as the state with the lowest energy. 
This state consists of an anti-ferro ordering of the MnO$_{2}$ planes along the z-direction
according to Mn$^{3+}$O$_{2}$/Mn$^{4+}$O$_{2}$.
Such a state is also observed in Sr
doped LaMnO$_{3}$. In the calculations of Banach and Temmerman\cite{banach} this state was
obtained by Sr doping whilst keeping the volume constant. The present results demonstrate 
that the volume reduction is another factor inducing the charge disproportionated state whilst
Jahn-Teller distortions, globally or locally, are not necessary. 
However, this charge order can drive a structural distortion which is
still present in LaMnO$_{3}$ under pressure.\cite{loa,pinsard-gaudart}
Note that we have only studied the simplest of the charge ordered states and that
a more complex ordering might give rise to a further lowering of the energy.
Obviously this could be modelled by larger supercells which was not examined 
in this work.

Consideration of the Jahn-Teller distortion in this volume range would probably introduce
the possibility of orbital order, which would be
interesting to study as a function of pressure.
The occurence of a charge disproportionated Mn$^{3+}$/Mn$^{4+}$ region
as a function of pressure could lead to local Jahn-Teller distortions
and the disappearance of the Jahn-Teller distortion as a function
of pressure might therefore not happen uniformly.

When the ferromagnetic charge disproportionated state is entered the system becomes metallic. 
The polarisation is around 70\% in this charge disproportionated state.
This state persists through the next phase transition into the
tetravalent Mn$^{4+}$ at a volume of 200 \AA$^{3}$. 

The charge disproportionation at the surface might be different from the bulk.
Calculations were performed to model a LaMnO$_{3}$ surface by considering a supercell
consisting of four LaMnO$_{3}$ formula units and two "empty LaMnO$_{3}$" formula units.
The LaMnO$_{3}$ surface can either be terminated with LaO or MnO$_{2}$ planes. 
Both were calculated and for a MnO$_{2}$ surface the delocalisation 
of the e$_{g}$ electron took place.
This is surprising since one would expect on intuitive
grounds, due to band narrowing, a localisation at the surface to be preferable.
This is the case for example in $\alpha$-Ce where the surface consists of $\gamma$-Ce.\cite{olle}
That a delocalisation at the surface took place is also reflected in a 
reduction of the Mn surface magnetic moment by 6\%. 
The reason might be that the e$_{g}$ electron localised in the bulk
is of $3r^{2}-z^{2}$ symmetry character and this orbital points perpendicular
to the surface into the vacuum. Therefore it has less electrons to capture for
localisation. Changing the surface termination to LaO does not alter 
the valence on the Mn in the MnO$_{2}$ sub-surface, however it does
change the symmetry of the localised state from $d_{3r^{2}-z^{2}}$ 
to $d_{x^{2}-y^{2}}$.
   
\section {Summary and Outlook}

We have calculated the electronic total energy of LaMnO$_{3}$, in the cubic structure,
as a function of volume and for different magnetic and  charge ordered states.
We find that the Jahn-Teller distortion is not crucial to explain the insulating
anti-ferromagnetic ground state of LaMnO$_{3}$. A smooth localization-delocalization transition
is observed since the transition from Mn$^{3+}$ to Mn$^{4+}$ goes through the intermediary
of a mixture of Mn$^{3+}$/Mn$^{4+}$. An accurate p-V curve could be modelled via CPA 
(coherent-potential-approximation)
calculations where at each volume the total energy is minimized with respect to
the concentration of Mn$^{3+}$ and Mn$^{4+}$.\cite{martin}\\

\section*{Acknowledgements}

G. Banach was supported by the EU-funded Research Training  Network: "Computational Magnetoelectronics" 
(HPRN-CT-2000-00143).
He also gratefully acknowledges discussions with Drs. A. Haznar and R. Tyer.

\end{document}